%% file: 0_main.tex
\documentclass[sigconf]{acmart}
\usepackage{amsmath,amsfonts}
\usepackage{algorithmic}
\usepackage{algorithm}
\usepackage{array}
\usepackage[caption=false,font=normalsize,labelfont=sf,textfont=sf]{subfig}
\usepackage{textcomp}
\usepackage{stfloats}
\usepackage{url}
\usepackage{verbatim}
\usepackage{graphicx}
\usepackage{xcolor}
\usepackage{pifont}
\usepackage{enumitem,enumerate}

\usepackage{amssymb}
\usepackage{amsthm}
\usepackage[english]{babel}
\newtheorem{theorem}{Theorem}
\usepackage{hyperref}
\hypersetup{
	colorlinks=true,
	linkcolor=blue,
	filecolor=blue,
	citecolor = blue,      
	urlcolor=blue
}

\newcommand{\yh}[1]{{\color{blue}#1}}

\renewcommand\footnotetextcopyrightpermission[1]{} 
\begin{document}
	
\title{Probabilistic Compute-in-Memory Design For Efficient Markov Chain Monte Carlo Sampling}
	

\author{Yihan Fu$^{1,2}$,
	Daijing Shi$^{1,2}$,
	Anjunyi Fan$^{1,2}$,
	Wenshuo Yue$^{1,2}$,
	Yuchao Yang$^{1,2,3,4*}$,
	Ru Huang$^{1,2,*}$,
	Bonan Yan$^{1,2,*}$}
\affiliation{%
	\institution{$^1$Institute for Artificial Intelligence, Peking University, Beijing, China; $^2$Beijing Advanced Innovation Center for Integrated Circuits, School of Integrated Circuits, Peking University, Beijing, China; $^3$School of Electronic and Computer Engineering, Peking University, Shenzhen, China; $^4$Center for Brain Inspired Intelligence, Chinese Institute for Brain Research (CIBR), Beijing, China}
	\streetaddress{}
	\city{}
	\state{}
	\country{}
}
\email{{yuchaoyang, ruhuang, bonanyan}@pku.edu.cn}

\input{1_abstract}

\maketitle

\input{2_intro}

\input{3_pre}

\input{4_op}	

\input{5_hardware}

\input{6_feature}

\input{7_eval}

\input{8_conclusion}

\input{9_appendix}

\bibliographystyle{ACM-Reference-Format}
\bibliography{ref}

\vfill
	
\end{document}

%% file: 1_abstract.tex
\begin{abstract}

Markov chain Monte Carlo (MCMC) is a widely used sampling method in modern artificial intelligence and probabilistic computing systems.
It involves repetitive random number generations and thus often dominates the latency of probabilistic model computing.
Hence, we propose a compute-in-memory (CIM) based MCMC design as a hardware acceleration solution.
This work investigates SRAM bitcell stochasticity and proposes a novel ``pseudo-read'' operation, based on which we offer a block-wise random number generation circuit scheme for fast random number generation.
Moreover, this work proposes a novel multi-stage exclusive-OR gate (MSXOR) design method to generate strictly uniformly distributed random numbers.
The probability error deviating from a uniform distribution is suppressed under $10^{-5}$. 
Also, this work presents a novel in-memory copy circuit scheme to realize data copy inside a CIM sub-array, significantly reducing the use of R/W circuits for power saving. 
Evaluated in a commercial 28-nm process development kit, this CIM-based MCMC design generates 4-bit$\sim$32-bit samples with an energy efficiency of $0.53$~pJ/sample and high throughput of up to $166.7$M~samples/s.
Compared to conventional processors, the overall energy efficiency improves $5.41\times10^{11}$ to $2.33\times10^{12}$ times.

\end{abstract}

%% file: 2_intro.tex
\section{Introduction}
\begin{figure}[t]
\centering
\includegraphics[width=0.9\columnwidth]{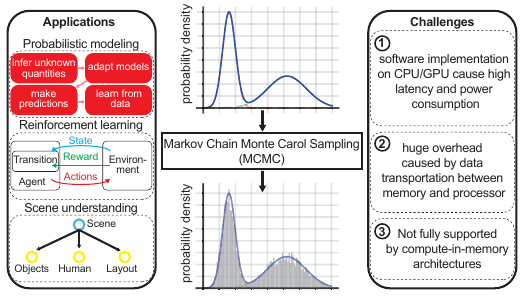}
\vspace{-10pt}
\caption{Applications and challenges of MCMC computing.}
\label{fig:mcmc1}
\vspace{-10pt}
\end{figure}

Markov chain Monte Carlo (MCMC) method is an essential sampling technique used in statistics and artificial intelligence (AI)~\cite{andrieu2003introduction}. 
It generates samples that closely approximate a given multivariate distribution, i.e. prior distribution~\cite{neal2000markov}. 
Unlike other typical sampling techniques such as rejection sampling, the MCMC method scales well with expanding state space and increasing dimensions. 
This algorithm provides a proper solution to those tasks featuring large datasets and high-dimensional models~\cite{karras2022overview}.
Specifically, MCMC sampling is widely used in probabilistic modeling~\cite{ghahramani2015probabilistic}, reinforcement learning~\cite{chung2020multi}, and scene understanding~\cite{chen2019holistic} (Fig.~\ref{fig:mcmc1}).

However, the execution of MCMC algorithm on central processing units (CPUs) or graphic processing units (GPUs) only generates $10^{2\sim 3}$ samples per second, which is far from the required sample generation speed for real-time probabilistic inference. 
For example, in the application of scene understanding~\cite{chen2019holistic++}, the system gets a parse graph as the optimal configuration of the 3D monitored scene at the beginning of one frame. 
The parse graph should be iteratively optimized by MCMC  sampling with simulated annealing based on posterior probability before the end of this frame.
But MCMC sampling based on GPU acceleration needs many seconds to generate enough samples for the optimization by frame, which apparently cannot meet the speed requirement, say, $33$~ms/frame ($30$ frame per second), for real-time scene understanding in robots.
The reason behind this is that every time a sample is generated in GPU, it needs to be transferred to the main memory. 
The frequent data transfer between processing units and memories causes long latency and high energy consumption, which dominates the overall computing process~\cite{liu2017communication,lao2020tfp,sebastian2020memory}.
This necessitates the invention of memory-centric hardware architectures to accelerate MCMC.

Prior works have explored MCMC sampling implementation on novel hardware platforms.
Cai et al.~\cite{cai2018vibnn} and Banerjee et al.~\cite{banerjee2019acmc} designed FPGA-based hardware accelerators which compromise sampling accuracy for gaining two to three orders of magnitude of performance-per-watt over CPU and GPU, but their work can only adapt MCMC sampling on Gaussian functions.
Shukla et al.~\cite{shukla2020mc} invented MC$^2$RAM using SRAM bitcells, multipliers, and analog-to-digital converters (ADCs) to achieve fast Gaussian mixture models (GMMs) sampling; however, it failed to thoroughly investigate the random sample precision and the influence of non-ideal practical conditions, e.g. process variations and thermal fluctuation.
Dalgaty et al.~\cite{dalgaty2021situ} exploited the variability of $16$k memristor nanoscale devices to implement random number generation and demonstrated low-power memristor-based MCMC for an exemplary reinforcement learning application, but it remains a proof-of-concept work at device research level that relies on the software platform and C++ standard random math package for the accept/reject check.

This work proposes a compute-in-memory (CIM) design to address these challenges to construct a high-performance MCMC sampler accelerator.
Featuring high sampling speed, expandable precision, and high parallelism, the proposed CIM-based MCMC design realizes a complete on-chip MCMC sampling iteration and has great potential as a standalone domain-specific core to be integrated into a system-on-a-chip (SoC).

The contribution of this work includes:
\begin{itemize}[leftmargin=*]
\item We investigate the stochasticity of SRAM bitcell and propose a novel operating principle of the ``pseudo-read'' operation, based on which we offer a block-wise random number generation (RNG) circuit scheme for fast random number (sample) generating speed over $10^6$ samples/s.

\item We propose a novel multi-stage exclusive-OR gate (MSXOR) design method to correct a biased probability (e.g. $~$40\%) to generate strict 50\% uniform distributed random numbers. The error probability achieves as low as $10^{-4}$\%. Further, we realize an efficient accept/reject check circuit module.

\item We propose a novel in-memory copy circuit scheme to realize data copy between different bitcells, which can significantly reduce the use of R/W circuits for energy- and power-saving purposes. 
\end{itemize}

To evaluate the proposed CIM design, we implement the proposed CIM-based MCMC design with a commercial process development kit (PDK) at 28~nm technology node with 6-transistor (6T) SRAM foundry bitcells. 
The proposed CIM-based MCMC design achieves high-accuracy sampling with an energy efficiency of $0.53$~pJ/sample and high throughput of up to $166.7$M~samples/s.

%% file: 3_pre.tex
\section{Preliminaries}\label{sec:2}

\subsection{Markov Chain Monte Carlo Sampling Algorithm}

\begin{algorithm}[b]
\caption{Metropolis-Hastings MCMC sampling}
\begin{algorithmic}
\STATE 
\STATE \hspace{0cm}$ \textbf{Initialise } x^{(0)} \subset X  $
\STATE \hspace{0cm}$ \textbf{For } \mathbf{i} =0 \textbf{ to } N-1 $ 
\STATE \hspace{0.5cm}Step 1: $\textbf{Sample } x^{*} \subset q(x^{*} | x^{(i)}) $
\STATE \hspace{0.5cm}Step 2: $\textbf{Sample } u \subset U[0,1] $
\STATE \hspace{0.5cm}Step 3: $\textbf{If } u< \alpha (x^{(i)},x^{*}) = min\{ {1,\frac{p(x^{*})q(x^{(i)}|x^{*})}{p(x^{(i)})q(x^{*}|x^{(i)})} } \} $
\STATE \hspace{2.2cm}$ x^{(i+1)} = x^{(*)} $
\STATE \hspace{1.6cm}$ \textbf{else} $
\STATE \hspace{2.2cm}$ x^{(i+1)} = x^{(i)} $
\end{algorithmic}
\label{alg:alg1}
\end{algorithm}

\begin{figure}[t]
	\centering
	\includegraphics[width=0.95\columnwidth]{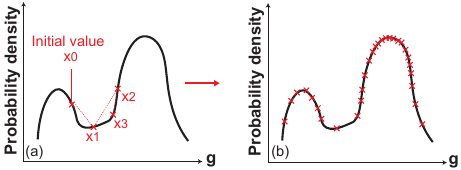}
	\vspace{-10pt}
	\caption{(a) Posterior distribution, initial value, and new samples drawn from MCMC sampling; (b) final accepted models to represent the distribution.}
	\label{fig:mcmcconcept}
	\vspace{-14pt}
\end{figure}

MCMC method is one of the fundamental building blocks of probabilistic inference to estimate the posterior probabilities and generate stochastic samples from designated probability distributions~\cite{ghahramani2015probabilistic}, especially those whose probability density function (PDF) is hard to be analytically expressed. 
It is an essential mathematical tool for solving integration and optimization problems in large dimensional spaces~\cite{andrieu2003introduction}. 
The core idea of MCMC is defining a Markov chain where the integral of state transfer probabilities should form a stationary distribution.
In other words, if the chain runs long enough time, the distribution of states will be proportional to the desired distribution.

Metropolis-Hastings (MH) MCMC sampling (Algorithm~\ref{alg:alg1}) is the most common approach in probabilistic computing~\cite{yildirim2012bayesian}. 
MH-MCMC is mostly for drawing samples directly from a prior distribution $p(x)$ toward the posterior distribution.
Fig.~\ref{fig:mcmcconcept} illustrates an MCMC sampling process.
Here, $x_0$ is the initial value as the first sample of the Markov Chain, randomly chosen in the set $X$ of all possible sampling values.
Then we randomly draw new samples through the transfer distribution $q(x)$ as new candidates for the following sample.
This Markov chain extends smoothly step by step in this way.
Finally, we can obtain a group of samples that can represent the stable distribution $p(x)$.
In this way, by defining the Markov state transfer probabilities, the MCMC method generates samples of the desired PDF $p(x)$.

An MH step of invariant distribution $p(x)$ and proposal distribution $q(x^{*}|x)$ involves sampling a candidate value $x^{*}$ (i.e. generating random number) given the current value $x$ according to $q(x^{*}|x)$.
Then through the calculation of acceptance probability, the Markov chain decides whether move to $x^{*}$ or remain at $x$. 
The most important part of this process is the generation of $x^{(i+1)}$.
And the value of $x^{*}$, which may become the value of $x^{(i+1)} $ after the calculation of accept ratio, is a random number generated from the previous value $x^{(i)}$ and the transfer function $q(x)$.

An important concept is the ``burn-in''~\cite{plummer2006coda}. 
It refers to the process that allows the Markov chain some iterations to reach its equilibrium distribution.
The number of iterative cycles (each cycle generates one candidate sample) is called ``burn-in time'' (or ``burn-in period'').
Only the samples from the equilibrium distribution of the design Markov chain is meaningful, i.e. we can only use the samples generated after the burn-in time.
Ideal calculation of MCMC often uses an empirical burn-in time of $500$ or $1000$ cycles. 

During MCMC calculation, considering the frequent data transfer during iterative random number generation (RNG) and accept-reject check, it is necessary to explore the in-memory implementation of MCMC to significantly improve computing efficiency.

\subsection{SRAM-Based Compute-In-Memory Macro}

\begin{figure}[t]
\centering
\includegraphics[width=1\columnwidth]{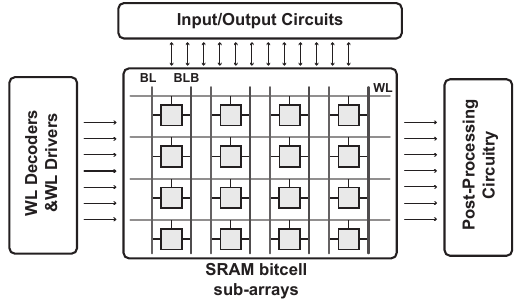}
\vspace{-10pt}
\caption{General circuit architecture of SRAM-based CIM.}
\label{fig:sram-cim}
\vspace{-10pt}
\end{figure}

CIM macro is a novel integrated circuit scheme that directly performs in-situ computational operations in memory.
As one of the promising solutions to overcome von Neumann bottleneck~\cite{RN2,RN3,RN292,moon2023parallel,pan2022mini},
this technology has been developed to perform energy-efficient computations in many AI applications~\cite{si202015,khwa201865nm,su202015,chih202116,kim20191,su202116,sinangil20207,zhang2022cp}.
The key idea of CIM is to fuse memory and process units.
SRAM is widely used in developing CIM macros for its fast read/write speed and good compatibility with standard CMOS fabrication processes, such as in $28$~nm, $7$~nm, and $5$~nm technology nodes~\cite{weste2015cmos}.
Fig.~\ref{fig:sram-cim} shows the basic structure of SRAM-based CIM macro~\cite{RN292}.
It consists of 6T-SRAM bitcell sub-arrays and peripheral circuits, including addressing modules, read/write circuits (i.e. sense amplifiers and write drivers), and computing circuits. 
With the assistance of the peripheral circuits, the bitcell sub-array can achieve the functions of storage and computation in independent working modes.
By offloading computation from centralized processing units, CIM dramatically eliminates the frequent data transportation between memory and processing units and thus forms an energy-efficient computing system~\cite{yan2019resistive}.
Even though CIM technology has been considered to generate random samples~\cite{dalgaty2021situ}, it has been rarely investigated in implementing full MCMC accelerator hardware with CIM technology.

%% file: 4_op.tex
\section{Operating Principles}\label{sec:3}
We propose to exploit the inherent stochastic characteristics of SRAM to develop novel peripheral circuits to realize MCMC hardware acceleration.
Before diving into the detailed circuit design, this section provides the basic principles used in the proposed CIM-based MCMC design.
This work leverages the inherent stochasticity of the SRAM bitcell as the source of  randomness.
Subsequently, based on this mechanism, the computing process of the CIM-based MCMC design is described in this section.

\subsection{Inherent Stochastic Characteristics of SRAM Bitcell}\label{sec:3.1}

\begin{figure}[t]
\centering
\includegraphics[width=1\columnwidth]{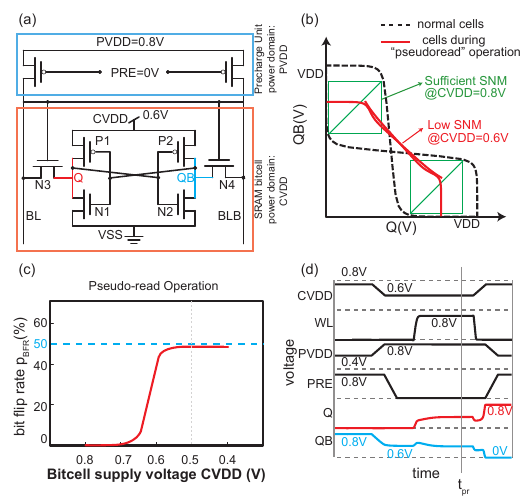}
\vspace{-10pt}
\caption{(a) ``Pseudo-read'' operation for an SRAM bitcell; (b) VTC of normal SRAM bitcells and the bitcells during ``pseudo-read'' operation; (c) bit flip rate of SRAM bitcells; (d) working condition and timing diagram for pseudo-read operation.}
\label{fig:concept}
\vspace{-16pt}
\end{figure}

Data stored in SRAM bitcells can be disturbed by thermal noise.
This work would like to leverage this as the source of controllable randomness.
According to the SRAM theory in~\cite{weste2015cmos}, each SRAM bitcell (illustrated in Fig.~\ref{fig:concept}(a)) has its own data retention voltage (DRV), which refers to the minimum supply voltage for an SRAM bitcell to preserve the datum bit resilient to the noise perturbation~\cite{qin2004sram}.
One common indicator for the stability of memory bitcells is the voltage transfer curve (VTC), usually called ``butterfly diagram'', as shown in Fig.~\ref{fig:concept}(b).
To quantify VTCs, we simulate with a commercial $28$~nm technology PDK and foundry 6T SRAM bitcells.
The standard bitcell supply voltage CVDD is $0.8$~V.
As Fig.~\ref{fig:concept}(b) shows, the edge length of the largest squares inscribed in the ``eyes'' of the butterfly curve are defined as static-noise margin (SNM)~\cite{calhoun2006static}. 
SNM quantifies the voltage noise necessary to flip the stored datum in a bitcell. 
When the supply voltage CVDD decreases, the two sides of VTC get close to each other, resulting in a sharp SNM decrease, implying that the bitcell resilience to thermal noise turns weaker.
Fig.~\ref{fig:concept}(c) shows the bit flip rate (BFR)~\cite{kagiyama2012bit,yamaoka201520k} throughout this process depending on the bitcell supply voltage CVDD.
SNM shrinks as the bitcell datum gain increased chances to flip randomly inflicted by  thermal noise.

However, DRV can be very small owing to the back-to-back connection of the two inverters in the SRAM bitcell.
When CVDD is around DRV, small changes of CVDD may cause rapid BFR fluctuation due to the circuit nonlinearity.
In order to gain better control of the SRAM bit flipping, we propose to define a novel operation, called ``pseudo-read''. 
It is an operation similar to SRAM read but it destroys the original datum bit stored in the SRAM bitcell and generates a random ``0'' or ``1''. 
Fig.~\ref{fig:concept}(d) shows the working condition and timing diagram for the bitcell pseudo-read.
During pseudo-read, we apply $0.8$~V to BL and BLB through the bitline conditioning transistors, with a lowered CVDD as the bitcell supply.
Then WL gets asserted and later deasserted to trigger the random bit flip phenomenon.
e.g. at the time of $t_{pr}$ in Fig.~\ref{fig:concept}(d), the operating configurations are shown in Fig.~\ref{fig:concept}(a).
During the pseudo-read, BL and BLB simultaneously keep high to accelerate the SNM shrinking.
We observe that BFR is around $45\%$ when CVDD lowers to $0.5$~V with this proposed pseudo-read operation\footnote{In reality, the value of the lowered CVDD is chosen according to the measured bit flip rate.}.
This is adequate as the source of randomness in the proposed MCMC sampling circuits. 

\subsection{Computing Steps of CIM-Based MCMC design}

\begin{figure}[t]
\centering
\includegraphics[width=1\columnwidth]{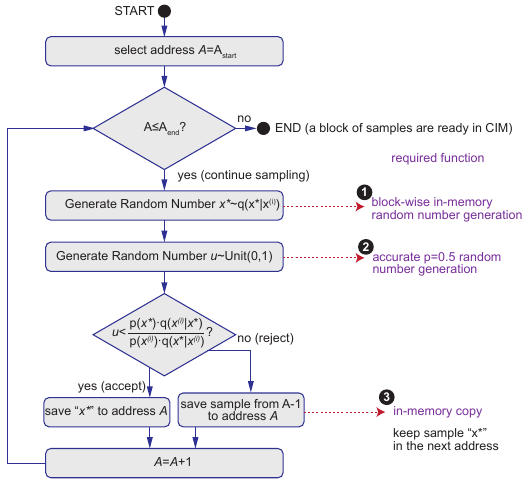}
\vspace{-10pt}
\caption{Block diagram for CIM-based MCMC method.}
\label{fig:opflow}
\vspace{-10pt}
\end{figure}

Fig.~\ref{fig:opflow} shows the flow chart of generating stochastic samples with the proposed CIM-based MCMC design.
The output of the MCMC is a set of samples $\{x_0,x_1,\cdots,x_m\}$ with required PDF $p(x)$, which are placed in the memory block from the address $A_{start}$ till $A_{end}$.
It consists of three major steps: for each memory address, (a) generate samples $x$ with the transfer PDF $q(x)$; (b) generate one sample $u$ from an accurate $[0,1]$ uniform distribution; (c) check the accept/reject condition: if accepted, the sample should be copied to a new address for results storage.

Here we elaborate on this process with an example of $4$-bit samples.
Each sample (number) is a value between $0$ ($0000$) and $15$ ($1111$). 
In this scale, the initial value $x^{(0)}$ is written into $4$ SRAM bitcells, each bitcell storing $1$ bit.
The sampled value $x^{*}$ follows the random distribution of $x^{*} \subset q(x^{*} | x^{(i)})$, and this process is achievable with pseudo-read operation explained in Sec.~\ref{sec:3.1}.

\begin{figure}[t]
\centering
\includegraphics[width=0.9\columnwidth]{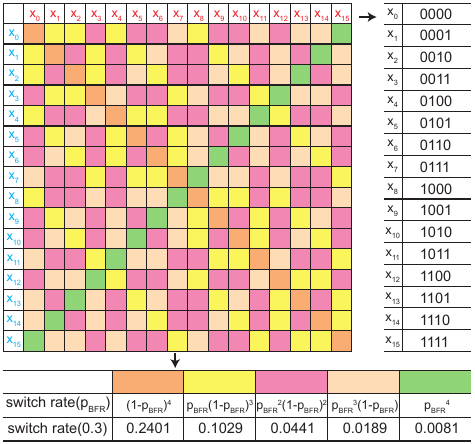}
\vspace{-10pt}
\caption{Probabilistic transfer matrix of 4-bit data.}
\label{fig:matrix}
\vspace{-10pt}
\end{figure}

For example, the stored value of 4 SRAM cells is ``$0$'' ($0000$).
After the pseudo-read, according to the curve of BFR (Fig.~\ref{fig:concept}(c)), these bitcells have a $45\%$ chance of generating ``$1$''.
We hereby obtain a sampled 4-bit random number $x^{*}$.
Then, the next step is to calculate the accept ratio $\alpha$.
Fig.~\ref{fig:matrix} shows the entire transfer matrix of a $4$-bit word. 
Given a certain CVDD for pseudo-read operation, we can measure and plot the transfer matrix $\mathbf{q}$, where we can look up for the values $q(x^{(i)}|x^{*})$ and $q(x^{*}|x^{(i)}$.
It is evident to find that this matrix is a diagonal matrix, with $q(i,j) \equiv q(j,i), for\ i,j \subset [0,15]$.
This transfer matrix $\mathbf{q}$ is bound to the condition of pseudo-read, and it is thereby a constant when calculating the accept ratio  $\alpha=\frac{p(x^{*})q(x^{(i)}|x^{*})}{p(x^{(i)})q(x^{*}|x^{(i)})}$.
Subsequently, $\alpha$ is simplified to $\frac{p(x^{*})}{p(x^{(i)})}$.

Given the required distribution $p(x)$, this value (``accept ratio'') can be easily calculated through peripheral digital logic circuits.
With an $[0,1]$ uniformly distributed random number $u$ generated, the comparison between $u$ and the calculated accept ratio decides whether the sample $x^*$ is qualified as the MCMC output sample.
According to Step 3 in Algorithm~\ref{alg:alg1}, the generated $x^*$, if accepted, will be copied to the next $4$ bitcells for subsequent sampling operations.
Otherwise, bitcells that store the unaccepted value will be overwritten the previous random value, and then this value will be copied to the next bitcells. 
In this way, CIM-based MCMC design fulfills the MCMC computing./l

%% file: 5_hardware.tex
 \section{CIM-based MCMC design Hardware Design}\label{sec:4}

\begin{figure*}[t]
\centering
\includegraphics[width=0.9\linewidth]{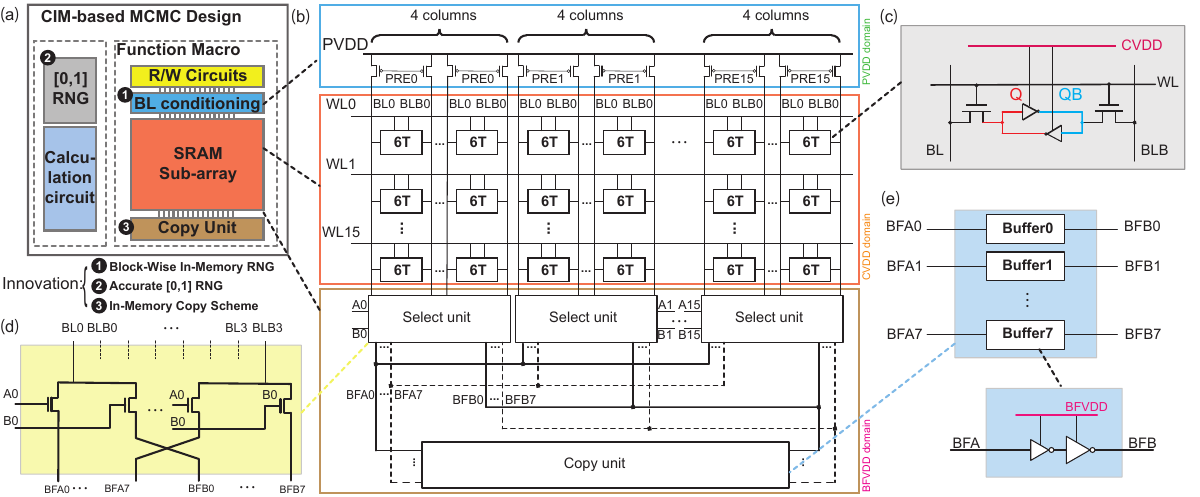}
\vspace{-10pt}
\caption{(a) Architecture of CIM-based MCMC design; (b) sub-array and peripheral circuits; (c) standard 6T-SRAM bitcells; (d) select unit; (e) copy unit.}
\label{fig:arc}
\vspace{-10pt}
\end{figure*}

\begin{figure}[t]
\centering
\includegraphics[width=1\columnwidth]{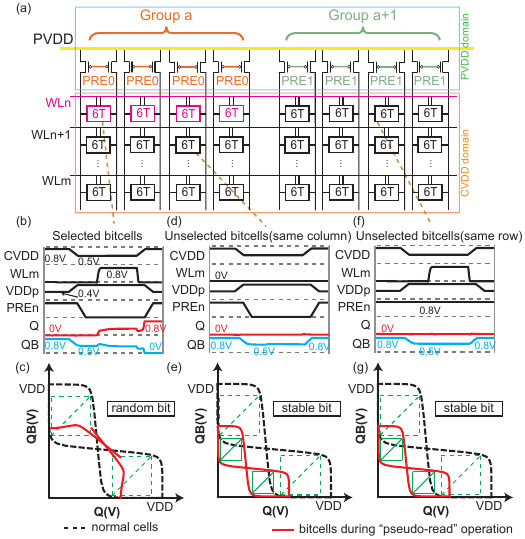}
\vspace{-10pt}
\caption{(a) Pseudo-read operation in a bitcell sub-array; (b) working condition and timing diagram of the selected bitcells; (c) VTC of the selected bitcells and unselected bitcells; (d) working condition and timing diagram of the unselected bitcells in the same column of the selected bitcells; (e) VTC of the unselected bitcells in the same column of the selected bitcells; (f) working condition and timing diagram of the unselected bitcells in the same row of the selected bitcells. (g) VTC of the unselected bitcells in the same row of the selected bitcells.}
\label{fig:RNG}
 \vspace{-10pt}
\end{figure}

As described in Sec.~\ref{sec:3}, CIM-based MCMC computation requires three major in-memory operators: \textit{in situ} in-memory RNG ($x^*$ generator), accurate $[0,1]$ RNG ($u$ generator) and in-memory copy.
This section elaborates on the circuit scheme to realize these three operations and form a complete CIM-based MCMC design.
CIM-based MCMC design performs on-chip MCMC sampling and storing the generated samples.
The entire MCMC processing happens locally inside the CIM-based MCMC design macro circuits, which eliminates the need for frequent data transfer between the processing units and memories in the conventional software realization on CPU/GPU platforms and thus significantly reduces the latency and energy consumption.
Fig.~\ref{fig:arc}(a) shows the overall architecture of the CIM-based MCMC design.
It consists of an SRAM sub-array\footnote{The SRAM sub-array is defined to consist  of SRAM bitcells, row/column decoders, and multiplexers. The latter two are not drawn here because of their conventional circuit structure.}, read/write (R/W) circuits, BL conditioning circuits(\ding{202}), in-memory copy units (\ding{203}), and accurate $[0,1]$ RNG module (\ding{204}).
Note that the former 4 modules form a function macro.
The accurate $[0,1]$ RNG module (\ding{204}) is a standalone module independent of the function macro.

This function macro can work in three modes:
\begin{itemize}[leftmargin=*]
    \item[$\circ$] \emph{Memory mode} works the same as traditional SRAM does. The functions can write (sense) data into (from) bitcells. In this mode, the SRAM sub-array, BL conditioning circuits, and R/W circuits work; all the other circuits are turned off.  
    \item[$\circ$] \emph{Block-wise RNG mode} generates random numbers into blocks. In this mode, only SRAM sub-array and BL conditioning circuits are turned on.
    \item[$\circ$] \emph{CIM copy mode} realizes copying data from one part to another part. In this mode, only the SRAM sub-array and the in-memory copy unit are turned on.
\end{itemize}

Next, we will detail how the proposed CIM-based design performs MCMC computing.

\subsection{Block-Wise In-Memory Random Number Generation}\label{sec:4.1}
Practical applications call for vast amounts of random samples before the accept/reject check.
According to Fig.~\ref{fig:opflow}, a large number of random numbers should be generated and immediately buffered in the memory for further MCMC processing steps. 
In order to achieve this, we design the 6T SRAM bitcell sub-array and block-wise in-memory RNG assisting circuit, as illustrated in Fig.~\ref{fig:arc}(b).
The entire function macro enters block-wise RNG mode.
As one realistic setup, every 4 columns of bitcells share one BL conditioning circuit (the precharge unit).
The power supply of the BL conditioning circuits is connected to PVDD and the bitcell sub-array connects to the supply voltage of CVDD.
In the block-wise RNG mode, pseudo-read\footnote{``Pseudo-read'' operation is defined in Sec.~\ref{sec:3.1}} is conducted for a block of the sub-array, as shown in Fig.~\ref{fig:RNG}(a).
The precharge unit of the selected cells is activated and the supply voltage connected to it is raised to $0.8$~V with PVDD=$0.8$~V, maintaining high voltage on both BL/BLB at the same time.
CVDD is reduced to $0.5$~V.
Then, the WL of the selected cells is activated.
As a result, the SNM of these bitcells are reduced purposely (as shown in Fig.~\ref{fig:RNG}(b) and (c)), and their values are destroyed.
For the bitcells in the same column (Fig.~\ref{fig:RNG}(d) and (e)),  since the WLs are not activated, their SNM remains high and the value stored is not destroyed~\cite{yamaoka201520k}.
For bitcells in the selected row (Fig.~\ref{fig:RNG}(f) and (g)), despite their corresponding WL activated, precharge units connected to them remain turned off.
The setting of a separate precharge unit makes it possible for the macro to perform "Random number generate" on specifically chosen cells, guaranteeing that the generation of new samples will not affect previous data that are stored in unselected blocks in the sub-array.

\subsection{Accurate \texorpdfstring{$[0,1]$}{[0,1]} Random Number Generation}
\label{sec:4.2}

\begin{figure*}[t]
	\centering
	\includegraphics[width=1\linewidth]{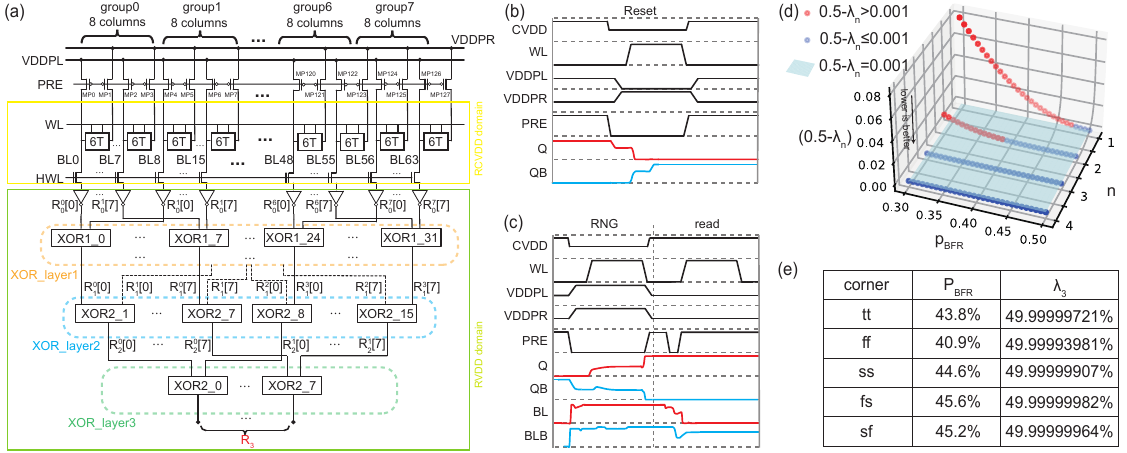}
	\caption{(a) Proposed accurate $[0,1]$ RNG architecture; (b) working condition and timing diagram of the reset operation; (c) working condition and timing diagram of RNG and read operation; (d) $\lambda$ of different bit flip rate and number of XOR$\_$stages; (e) $\lambda_{3}$ of SRAM bitcells under different corner simulations. }
	\label{fig:01}
\end{figure*}

After the candidate samples for the desired Markov chain are obtained from the block-wise RNG, whether they can be accepted as the output samples depends on the accept/reject check step.
That is, compare a random number $u$ and the expression $\alpha (x^{(i)},x^{*})$ in Algorithm~\ref{alg:alg1}.
This requires the random number $u$ to be uniformly distributed between $[0,1]$.

The challenge of realizing this accurate $[0,1]$ RNG is two-fold:
(a) software~\cite{phillips2011pseudo} or digital-logic-based pseudo-random number generation algorithms potentially cause large area and energy overhead~\cite{rezk2019reconfigurable};
(b) since the SRAM bitcell BFR $p_{BFR}$ under pseudo-read operation is lower than $50\%$  as shown in Fig.~\ref{fig:concept}(c), direct use of pseudo-read cannot guarantee that the generated random number $u$ is uniformly distributed\footnote{For a multi-bit integer $u$, uniform distribution can be interpreted as that each binary bit of $u$ has a strict $50\%$ probability of being ``1'' or ``0''.}, which potentially bring adverse effects on the quality of MCMC sampling.
Therefore, we propose a novel low-cost multi-stage XOR gate circuit scheme (MSXOR) to construct the accurate $[0,1]$ RNG.

Fig.~\ref{fig:01}(a) depicts the circuit schematic diagram of the proposed accurate $[0,1]$ RNG.
Accurate $[0,1]$ RNG includes a small-scale 6T SRAM bitcell sub-array and a group of multi-stage XOR gates.
This small-scale SRAM bitcell sub-array has a similar structure to the sub-array used block-wise RNG in Sec.~\ref{sec:4.1}; yet the former has no R/W circuits.
Besides, in an accurate $[0,1]$ RNG module, the voltage VDDP is divided into two voltages VDDPL and VDDPR to respectively control BL and BLB.

This scheme requires two steps to generate a multi-bit accurate $[0,1]$ random number:

\emph{Step 1:} Flush the employed SRAM bitcells to zeros (``reset''). This step can be performed through the voltage difference of VDDPL and VDDPR, as illustrated in Fig.~\ref{fig:01}(b).
VDDPL and VDDPR are both controlled by the transistors MP0$\sim$MP127 with the same control signal PRE.
In the reset operation, the supply voltage of the SRAM bitcells is firstly reduced to a relatively low value (0.5V), then the supply pair (VDDPL, VDDPR) is set to ($0$~V, $0.8$~V).
Then the MP0$\sim$MP63 and WL are turned off.
Since VDDPR is higher than the bitcell supply voltage, the stored value of these bitcells is modified to ``$0$''. 

\emph{Step 2:} Perform the pseudo-read RNG operation on these bitcells, which is similar to the step described in Sec.~\ref{sec:4.1}.
Considering the separate VDDPL and VDDPR controlling BL/BLB, both VDDPL and VDDPR should be raised to $0.8$~V in the pseudo-read steps.
The waveform of the control signals is depicted in Fig.~\ref{fig:01}(c).

Next, we will show why this scheme with the two-step operation yields samples of uniform distribution.
First, we define the probability of a single bit to be ``1'' as $\lambda$.
The multiple stages of XOR gates help optimize $\lambda$ of the generated samples to $0.5$ and equivalently the aggregated multi-bit results achieve uniform distribution.
In the circuit scheme, $64$ bitcells are divided into $8$ groups, i.e. each group representing an $8$-bit ``raw'' random number $R_0^l[7:0]$ ($l \in [0,7]$).
Note that the probability of each bit in $R_0^l[7:0]$ as ``1'' is $\lambda_0=p_{BFR}$ because it is right out of reset and pseudo-read operation.
The subscripts of $R_0^l$ and $\lambda_0$ mean it is the first stage.

Then it feeds $R_0^0, R_0^1, \cdots, R_0^7$ into a stage of $64$ XOR gates.
Each XOR gate converts two input bits into one output bit.
Therefore, at this stage, the XOR gates output $R_1^0$, $R_1^1$, $R_1^2$, $R_1^3$.
As illustrated in Fig.~\ref{fig:01}(d), both the two inputs at this stage have a probability of $\lambda_0$ being ``1'' and a probability of $1-\lambda_0$ being ``0''.
The output bit of the XOR gate hereby has a probability of $\lambda_1=2\lambda_0(1-\lambda_0)$ to be ``1'' (and a probability of $(1-\lambda_1)$ being ``0''). 

Then $R_1^0,R_1^1,R_1^2,R_1^3$ successively pass stage $2$ ($32$-bit XOR gates) and stage $3$ ($16$-bit XOR gates).
Finally, this module yields 1 group of samples $R_3[7:0]$ as the final output of accurate $[0,1]$ RNG in $8-bit$.
Take $p_{BFR}=0.4$ as an example, $\lambda_3=0.49999872$. 
The detailed proof of $\lim\limits_{n\rightarrow\infty}\lambda_n=0.5$ can be found in Appendix~\ref{sec:appx1}.
Fig.~\ref{fig:01}(d) shows the quantitative analysis result of the distance from $\lambda=0.5$, i.e. $(0.5-\lambda_n)$, given the value of $p_{BFR}$ and the number of XOR gate stages $n$.
It reveals that when $p_{BFR}\ge 0.4$ (corresponding to the case of $CVDD$ is disturbed from $0.5$V to $0.6$V), $3$-stage XOR-gates ($n=3$) is adequate.

\textbf{Discussion on the impact of process variation}: as the $[0,1]$ RNG circuit is composed of a limited number of SRAM bitcells, process variation of these bitcells is an adverse factor that potentially affects the accuracy of uniform distribution. 
 To address this issue, this work carries out the corner simulation for accurate $[0,1]$ RNG circuit. 
The simulation uses CVDD=$0.5$V in pseudo-read as the controlled variable.
Fig.~\ref{fig:01}(e) summarizes the corresponding results.
It reveals that $p_{BFR}$ fluctuates in different design corners as influenced by the process variation. 
This fluctuation arises because the operating period may not be appropriate for the SRAM bitcells to reach an intermediate voltage level during pseudo-read. 
Our simulation shows that extending the latency of pseudo-read operation can ameliorate this problem.
Fig.~\ref{fig:01}(e) shows that the proposed $3$-stage XOR gates yield $\lambda_3\ge 0.4999993981$, which is adequately accurate for the uniform distribution RNG and thus passes the corner test.

After a candidate sample $x$ (from the block-wise in-memory RNG) and a uniformly distributed random number $u$ (from accurate $[0,1]$ RNG) are prepared, the accept ratio calculation is then performed with digital arithmetic units. First, convert the random number $u$ into the range of $[0,1]$. Since $R_3$ is an uniformly distributed $8$-bit number between 0(\texttt{0x00}) and 255(\texttt{0xff}). The desired random number can be calculated through $u/256$. 
Then calculate the expression $\frac{p(x^{*})q(x^{(i)}|x^{*})}{p(x^{(i)})q(x^{*}|x^{(i)})} $ and compare the value with $u$.
To Reduce the complexity of the program, we can change the algorithm from comparing the size of fractions to comparing representation $u*{p(x^{*})}$ and ${p(x^{(i)})}$. If ${p(x^{(i)})}>u*{p(x^{*})}$, the new sample should accepted.
In this case, an in-memory copy operation should be carried out to store the accepted sample. 

\subsection{In-Memory Copy Scheme}
After obtaining an accepted sample, it should be written to another nearby address for further use.
A normal ``read-and-write'' operation through SRAM R/W circuits to accomplish this is straightforward but timing-consuming.
Because the value to be written back to the array is exactly stored in the previous group of bitcells, a method of copying data between neighboring bitcells could be a better solution for reducing read-write latency and power consumption. 

\begin{figure}[t]
\centering
\includegraphics[width=1\columnwidth]{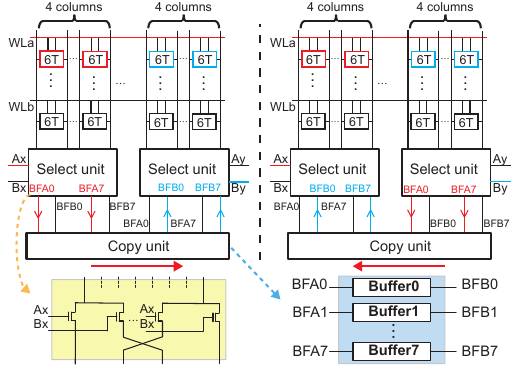}
\vspace{-10pt}
\caption{In-memory copy scheme.}
\label{fig:copy}
\vspace{-14pt}
\end{figure}

To achieve this, we propose an in-memory copy scheme, as shown in Fig.~\ref{fig:copy}. 
Every 4 columns of bitcells in the sub-array form a group with two control signals $A$ and $B$. 
Note the 4-column configuration follows Fig.~\ref{fig:arc}(b).
Every compartment of the macro shares 8 buses, which are corresponding to BLs and BLBs from every group of bitcells. 
Every bus has one buffer\footnote{Here the buffer refers to a circuit buffer, i.e. even number of inverters to increase the drivability.}, whose supply voltage remains high ($0.8$~V). 
Since the buffer is unidirectional, every BL (BLB) is connected to both sides of the corresponding buffer with signal $A$ controlling the switch on the input port and $B$ controlling the switch on the output port. 
To copy the value from bitcells from columns $[a_0,b_0,c_0,d_0]$ to columns $[a_1,b_1,c_1,d_1]$ in the same row, the supply voltage of the sub-array is reduced to $0.5$~V. 
Then the WL of this row is activated and the control signal $A0$ corresponding to $[a_0,b_0,c_0,d_0]$ on the buffer input port and $B1$ corresponding to $[a_1,b_1,c_1,d_1]$ on the buffers' output port are raised. 
In this way, the values of the bitcells in $[a_0,b_0,c_0,d_0]$ are transferred to the voltage on BLs and BLBs and enlarged to $0.8$~V through the associated buffers.
They hereby are written into the bitcells of $[a_1,b_1,c_1,d_1]$ with a relatively low supply voltage. 
After a short writing latency ($<1$~ns in our experimental configuration), the WL turns off and the supply voltage of the whole array restores $0.8$~V. 
In this way, an efficient in-memory copy operation is realized. 

%% file: 6_feature.tex
\section{Enhancement for CIM-based MCMC Design}\label{sec:5}
The proposed CIM-based MCMC design is expandable in the precision of generated samples and scalable for high parallelism.

\begin{figure*}[t]
	\centering
	\includegraphics[width=0.9\linewidth]{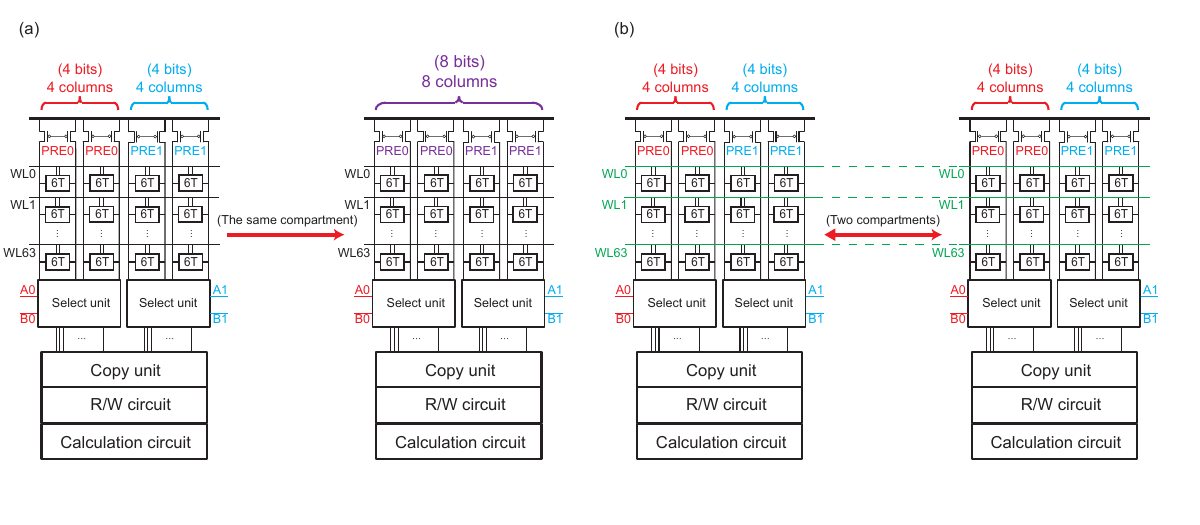}
	\vspace{-14pt}
	\caption{(a) Expandable precision scheme; (b) scalable parallelism scheme.}
	\label{fig:ep}
	\vspace{-14pt}
\end{figure*}

\subsection{Expandable Precision}
Earlier, when we describe the working principle of the proposed macro, we take the example of a word composed of $4$ bits.
Actually, the proposed CIM-based MCMC design supports higher precision as required in some applications~\cite{raftery1996implementing}.
In this way, we take a precision-expandable scheme to solve this problem. As shown in Fig.~\ref{fig:ep}(a), the array described in this paper is divided into several compartments, with each compartment composed of 4096 SRAM cells (64 rows \& 64 columns).
Since in the most basic case these bitcells are divided into 16 groups and each group consists of 4 columns, we can combine 2 groups of bitcells into a new group by keeping their control signal at the same pace with each other.
When performing the sampling, these control signals are turned on simultaneously, enabling outputting an 8-bit sample at a time.
Under 
\yh{R/W} operations and in-memory copy operations, we can apply the principle of separate transmission, operating on the first 4 columns and then the next 4 columns.
Through this method, we can conduct in-memory copy for the samples of up to 64 bits.

\subsection{Scalable Parallelism}

\begin{figure}[t]
	\centering
	\includegraphics[width=1\linewidth]{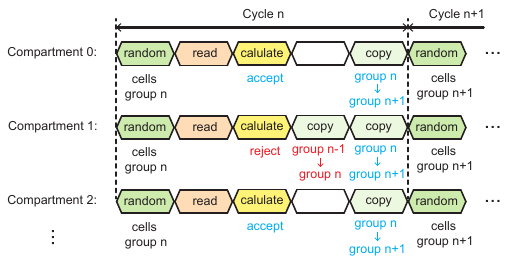}
	\vspace{-12pt}
	\caption{Operating timing diagram of this probabilistic macro.}
	\label{fig:pipeline}
	\vspace{-12pt}
\end{figure}

For computing, high parallelism is what researchers always pursue.
The proposed one macro can only perform MCMC sampling once at a time, failing to deliver adequate parallelism.
Therefore, we propose to improve the parallelism scheme by dividing the whole sub-array into different compartments.
Fig.~\ref{fig:ep}(b) depicts this scheme.
Each compartment is composed of 4096 SRAM bitcells (64 rows, 64 columns), copy circuit for the in-memory copy scheme, R/W circuit, and calculation circuits.
When we perform sampling, every compartment is thrown into operation.
The working timing diagram of a compartment-based macro is shown in Fig.~\ref{fig:pipeline}.
For every step (``random'' means block-wise RNG, ``read'' means reading out the generated sample and performing accurate $[0,1]$ RNG simultaneously, ``calculate'' means the calculation of accept/reject check, and ``copy'' means ), WLs of all compartments and corresponding control signals are turned on simultaneously so as to output one sample per compartment instead of per entire macro.
The acceptance probability calculation operation is performed at the same time for every compartment.
Of course, some samples will be accepted and others will be refused.
Therefore, two independent groups of in-memory copy circuits can solve this problem.
In one group of in-memory copy circuits, the compartments which obtain accepted samples do not need to modify themselves, as the remaining should copy the previous sample to the present address.
This operation can also be performed simultaneously, with WLs of the accepted compartments turned off and the remaining WLs turned on.
In the other group of in-memory copy circuits, all compartments are involved, with the value of bitcells of the present address copied to bitcells of the next address.
In this way, a group of newly sampled data is obtained and the Markov Chain keeps evolving~\cite{moon2023parallel}.

%% file: 7_eval.tex
\section{Evaluation and Discussion}\label{sec:6}
\subsection{Configuration for Verification \& Circuit Implementaiton}

\begin{figure}[b]
\vspace{-10pt}
\centering
\includegraphics[width=1\columnwidth]{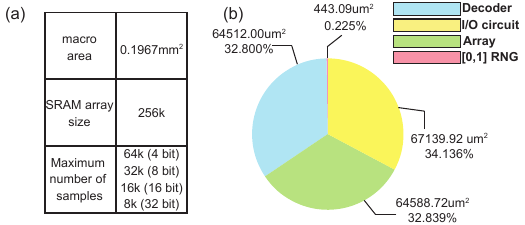}
\vspace{-10pt}
\caption{(a) Design parameter of the proposed macro; (b) the total area breakdown of this macro.}
\label{fig:area}
\end{figure}

The proposed CIM-based MCMC scheme is designed in the latest commercial $28$~nm CMOS process development kit.
The 6T SRAM bitcell is the foundry-developed bitcell.
The capacity of this macro is $256$~kb and the core area is $0.1967\ \mathrm{mm}^{2}$.
Fig.~\ref{fig:area}(a) lists the design parameter of this macro.

The area is estimated based on the layout of prior practical SRAM-based CIM design~\cite{RN292}.
Fig.~\ref{fig:area}(b) presents the pie diagram for the entire area breakdown.
Generally, R/W circuits dominate the area of this macro ($34.136\ \%$).
SRAM sub-array with select and copy units is the second largest part of this macro ($32.839\ \%$).
It is slightly larger than the WL decoders ($32.800\ \%$).
Since one random number generated from accurate $[0,1]$ RNG can be shared by the overall $64$ compartments. 
The proposed accurate $[0,1]$ RNG is very compact, only occupying 0.225$\%$ of the entire area.

\subsection{Function Verification}
\begin{figure*}[t]
\centering
\includegraphics[width=1\linewidth]{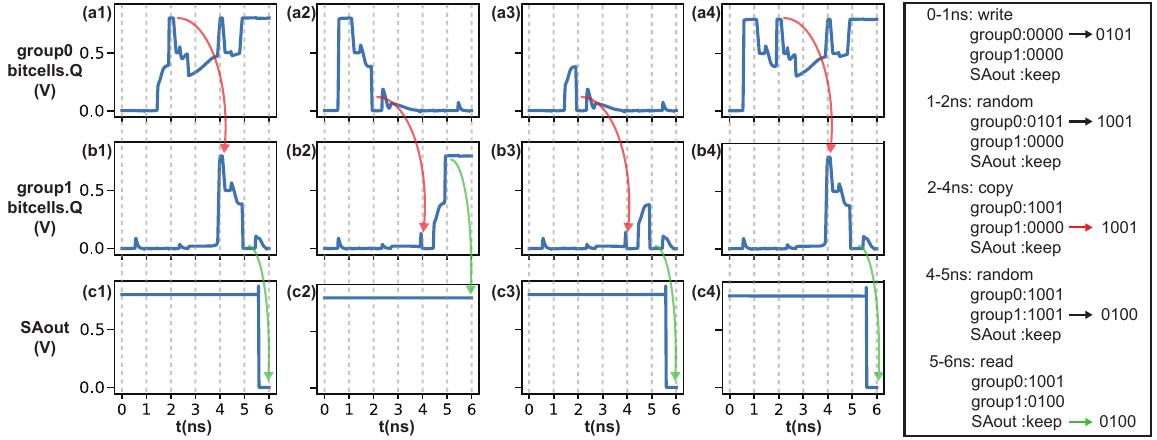}
\vspace{-10pt}
\caption{Function realization of the proposed macro. }
\vspace{-10pt}
\label{fig:func}
\end{figure*}
In this section, the function of the proposed macro is verified. 
Fig.~\ref{fig:func} shows the simulated timing diagram of SRAM bitcells as well as the corresponding output of R/W circuits.

Fig.~\ref{fig:func}(a1-a4) correspond to SRAM bitcells belonging to the first 4 columns, i.e. columns (0,1,2,3).
Fig.~\ref{fig:func}(b1-b4) correspond to the second 4 columns, i.e.columns (4,5,6,7).
Fig.~\ref{fig:func}(c1-c4)  corresponds to the output of the R/W circuit.
``Q'' refers to the positive terminal inside the 6T SRAM bitcell in Fig.~\ref{fig:arc}(b).

\begin{itemize}[leftmargin=*]
	\item From 0~ns to 1~ns (x-axis), the proposed macro goes through the ``write'' process, writing an exemplary value ``0101''into the 4 SRAM cells belonging to columns (0,1,2,3).
	\item From 1~ns to 2~ns, block-wise RNG is performed (marked with``random'' in Fig.~\ref{fig:func}).
	The originally stored data changes from ``0101'' to ``1001''.
	This new data ``1011'' serve as the first sample of the Markov chain.
	\item From 2~ns to 4~ns, the macro executes ``in-memory copy''.
	The data ``1001'' stored in columns (0,1,2,3) are copied to the bitcells of columns (4,5,6,7).
	\item From 4~ns to 5~ns, the macro executes block-wise RNG on the bitcells of columns (4,5,6,7), with their stored data switching from ``1001'' to ``0100''. 
	\item From 5~ns to 6~ns, the new data ``0100'' is read through the output circuit for the accept/reject check. 
\end{itemize}
In this way, the proposed macro gets a group of samples stored in the SRAM array following the desired distribution with the MCMC method.

\subsection{Thermal Impact}

\begin{figure}[t]
\centering
\includegraphics[width=1\columnwidth]{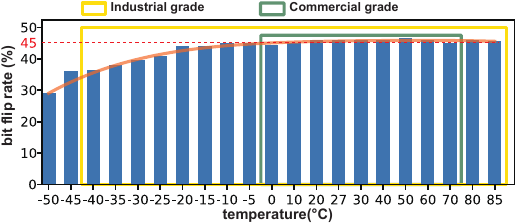}
\vspace{-10pt}
\caption{Bit flip rate under different temperatures.}
\label{fig:temp}
\vspace{-10pt}
\end{figure}

The operation of the proposed macro strongly depends on SRAM bitcell stochastic behavior~\cite{suresh2013analyzing}.
The stochasticity is strongly related to the ambient temperature.
Different temperatures may affect the working efficiency of the random generation.
Keeping the supply voltage $0.5$~V, we measure bit flip rate $p_{BFR}$ at different temperatures.
As shown in Fig.~\ref{fig:temp}, $p_{BFR}$ slightly increases as the temperature rises.
In the normal commercial-grade operating temperature range$0^\circ C-70 ^\circ C $, $p_{BFR}$ remains around $45\%$.
This implies the influence of the ambient temperature. 
For industrial grade ($-40^\circ C-85 ^\circ C$), this macro can work in most temperature ranges ($-20-85 ^\circ C $) with a stable bit flip rate.
When working at a lower temperature($-40--20 ^\circ C $), $p_{BFR}$ decreases as thermal noise reduces. 
According to the prior study in ~\cite{andrieu2003introduction}, the decrease in bit flip rate will only result in an extension of burn-in time to generate random samples without affecting the quality of obtained samples.
For the accurate $[0,1]$ RNG, according to Fig.~\ref{fig:01}(d) and (e), a slight decrease of $p_{BFR}$ is tolerable for maintaining the correct RNG function.

\subsection{Energy and Efficiency}

\begin{figure}[t]
\centering
\includegraphics[width=1\columnwidth]{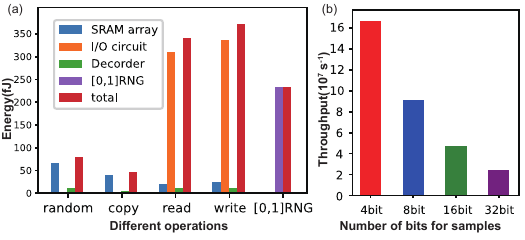}
\vspace{-10pt}
\caption{(a) Energy consumption and breakdown of different operations; (b) throughput of the proposed macro with different precision of samples.}
\label{fig:et}
\vspace{-10pt}
\end{figure}

Fig.~\ref{fig:et}(a) shows the energy consumption of the proposed macro under different operations.
When performing the block-wise RNG and in-memory copy operations, this macro consumes $79.1$~fJ and $47.5$~fJ for each $4$-bit sample, respectively.
Generally, the operations involving R/W circuits consume more energy than those completely inside the SRAM sub-array ($343.1$~fJ for read operation and $372.6$fJ for write operation).
This again confirms the necessity and importance of the in-memory copy scheme.
By avoiding excessive reuse of the 
R/W circuit in frequent data transportation,  the proposed macro significantly reduces energy consumption in accelerating MCMC computation.
The energy consumption of $[0,1]$ RNG is $234.6$~fJ for each 8-bit sample.

Taking everything into account, the energy cost of this entire macro is $0.5065$~pJ/accepted sample, for the sampling operation whose result is a directly accepted sample.
However, if the obtained sample is rejected by the calculation circuit, the energy cost of this operation will deteriorate as we should conduct an extra in-memory copy operation to rewrite the value of the previous sample to this rejected sample.
In this case, the energy cost is $0.5547$~pJ/sample.
The sampling accept ratio typically remains between $30\%$ and $40\%$ based on different prior distributions. Generally, considering the overall acceptance, the energy efficiency of the proposed macro will range from $0.5331$~pJ/sample to $0.5402$~pJ/sample.

\subsection{Throughput}
Fig.~\ref{fig:et}(b) illustrates the throughput of the proposed macro with different precision of samples.
As the number of bits in the sample increases, the overall throughput shows a continuous decrease but still maintains above $10^{7}$~samples/s.
It proves the proposed macro keeps a high sampling speed.
It is notable that as the number of bits doubles, the overall throughput of this macro does not decrease exactly by half.
In fact, it is slower.
This is because when we perform sampling operations with different precision, in-memory copy, R/W operations are performed step by step with 4 SRAM bitcells. As the number of bits doubles the time consumed in these operations will also double.
But the in-memory random operation can be performed simultaneously for any number of bits, as the selected SRAM cells are always in the same row and their WL is opened consistently. 
By adjusting pre-charge units of different columns, we can realize the generation of a complete sample in one sample operation, eliminating the time consumed to proceed in steps. 
Thus generally it takes less than twice the time to generate a sample of double number of bits. 
The overall throughput will accordingly decrease slower than decline by half.

\subsection{Comparison to MCMC Benchmarks on CPU\&GPU}

\begin{figure}[t]
\centering
\includegraphics[width=1\linewidth]{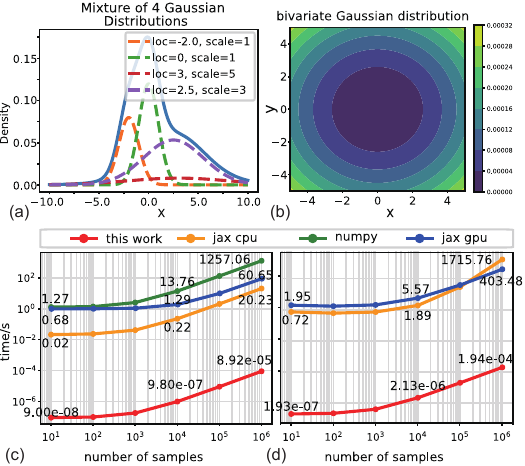}
\vspace{-10pt}
\caption{(a) Mixture of 4 Gaussian distributions; (b) two-dimensional heat map for a bivariate MGD example; (c) experimental result of the GMM case; (d) experimental result of the MGD case.}
\label{fig:018}
 \vspace{-10pt}
\end{figure}

In this part, we compare the performance of the proposed macro with CPU and GPU on different sampling tasks. 
We designate two probability distributions commonly used in artificial intelligence systems for MCMC, as illustrated in Fig.~\ref{fig:018}(a) and (b).
The first one is the Gaussian mixture model (GMM)~\cite{reynolds2009gaussian} in Fig.~\ref{fig:018}(a), a probabilistic model widely used for representing the distribution of data points in a dataset. 
It assumes that the dataset is generated from a mixture of several Gaussian distributions, each characterized by its mean, covariance, and weight. 
The mean represents the center of this distribution; the covariance captures its shape and orientation; and the weight indicates the contribution of each component to the overall mixture. 
Here we use a mixture of $4$ Gaussian distributions for GMM.
The other one is multivariate Gaussian distribution (MGD)~\cite{do2008multivariate}, also known as a multivariate normal distribution. 
It is a probability distribution that generalizes the concept of the Gaussian distribution to multiple dimensions. 
Fig.~\ref{fig:018}(b) shows the two-dimensional heat map for a bivariate MGD example.
Darker colors represent lower probabilities and lighter colors represent higher probabilities. 
This distribution is widely used in the tasks such as pattern recognition~\cite{theodoridis2006pattern} and anomaly detection~\cite{chandola2009anomaly}. 

In the experiment, we used CPU, GPU, and simulation results of the proposed macro to test the sampling efficiency on these two distributions. 
For GMM, we performed sampling on the CPU with the NumPy~\cite{harris2020array} and its adapted algorithms. 
Additionally, we employ JAX library~\cite{bradbury2021jax} with just-in-time compilation and executed along for faster speed on both CPU and GPU.
The CPU hardware platform is Intel\textsuperscript{\textregistered} Xeon\textsuperscript{\textregistered} Gold 6230 CPU @2.10GHz with 220GB main memory.
The GPU hardware platform is NVIDIA\textsuperscript{\textregistered} GeForce\textsuperscript{\textregistered} RTX 3090, and the samples drawn from our probabilistic macro are fixed at 32 bits. 

The experimental result is shown in Fig.~\ref{fig:018}(c) (GMM case) and Fig.~\ref{fig:018}(d) (MGD case). 
The x-axis is the target number of sample generation using the MCMC algorithm and the y-axis is the time cost for completing the corresponding MCMC tasks.

For software algorithms, there are two points that need to be explained. 
The first is that the sampling speed of JAX of CPU is faster than GPU when performing sampling for GMM.
This is because GMM is a relatively simple distribution, and the number of samples is rather low. 
When performing large-scale highly parallel computations, GPUs have an advantage over CPUs, but this advantage does not serve totally for sampling GMM. 
In fact, when we perform sampling on MGD, the speed advantage of GPUs over CPUs becomes evident as the number of samples increases. 
The second is we did not use NumPy algorithms when performing GCD sampling.
This is because the NumPy library is generally used for handling relatively simple computations, and it has been proven to be challenging to use NumPy and related algorithms for sampling on MGD.

Fig.~\ref{fig:018}(c) compares the sampling speed of this macro with different algorithms on CPU and GPU on GMM. 
The sampling algorithm using the NumPy library on the CPU takes more than $1200$~s ($20$~min) to generate $10^6$ samples. 
The use of the JAX library significantly improves sampling efficiency~\cite{bradbury2021jax}, but the sampling time remains more than $10$~s on both CPU and GPU. 
Our proposed macro gives a boost on sampling throughput, with $10^6$ samples generated within $10^{-3}$~s, showing an acceleration of over $20,0000\times$.

Fig.~\ref{fig:018}(d) shows the sampling speed on a bivariate MGD. 
The sampling speed shows a noticeable decrease compared to GMM as the distribution grows more complex. 
As the number of samples increases, the advantage of GPU over CPU in data processing becomes evident. However, the JAX algorithm based on GPU still takes around $400$ seconds to generate $10^6$ samples. 
On this complex sampling task, the superiority of our chip is further demonstrated, as the proposed macro only takes around $2$~ms to draw $10^6$ samples.

The power efficiency of the sampling of the proposed macro is also significantly higher than GPU.
For GMM, the average power consumption of GPU is around $125$~W, compared to $0.157$~mW on our chip. 
For MGD, the average power consumption of GPU is around $170$~W, compared to $1.52\times10^{-4}$~W on our chip. 
Taking sampling speed and power efficiency into account, the overall energy cost of obtaining one sample of our chip is $5.41\times10^{11}$ to $2.33\times10^{12}$ times lower than that of GPU.

%% file: 8_conclusion.tex
\section{Conclusion}\label{sec:7}
This work proposes an SRAM-based compute-in-memory macro for Markov Chain Monte Carlo (MCMC) sampling.
Featuring high sampling speed, expandable precision, and high
parallelism, this CIM macro can realize the complete
MCMC sampling process, including on-chip sampling, random
number generating, accept ratio calculation, and in-memory data copy. Implemented in a standard 28-nm CMOS process,
this macro can achieve high energy efficiency of $0.53$~pJ/sample and high throughput of up to $166.7$M~samples/s. Generally, our chip has a power consumption that is 6 orders of magnitude lower than that of a GPU.

%% file: 9_appendix.tex
\appendix

\section{Appendix: Proof of \texorpdfstring{$\lim\limits_{n\rightarrow\infty}\lambda_n=0.5$}{limit of lambda is 0.5}}
\label{sec:appx1}
\begin{figure}[h]
	\vspace{-12pt}
	\centering
	\includegraphics[width=0.85\linewidth]{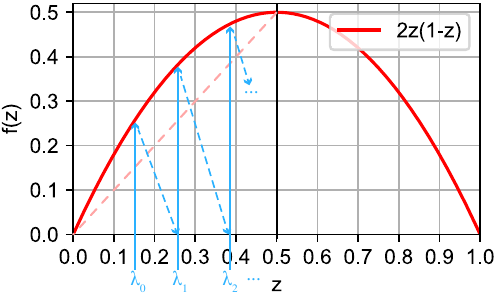}
	\vspace{-12pt}
	\caption{$f(z)=2z(1-z)$ and iteration of $\lambda_n$.}
	\label{fig:fz}
	\vspace{-12pt}
\end{figure}

The proposed multi-stage XOR gate scheme $\lambda_{i+1}=2\lambda_i(1-\lambda_i)$.
The function is abstracted as $f(z)=2z(1-z)$.
The curve of $f(z)$ is shown in Fig.~\ref{fig:fz}.

\begin{theorem}\label{theorem1}
	$\forall 0<z<0.5$,  $f(z)>z$.
\end{theorem}
\begin{proof}
 $\forall 0<z<0.5$, $f(z)-z=2z(0.5-z)$. $\because z>0$ and $0.5-z>0$, $\therefore f(z)>z$.
\end{proof}

\begin{theorem}
	$\forall 0<\lambda_0<0.5$,  $\lim\limits_{n\rightarrow\infty}\lambda_n=0.5$.
\end{theorem}
\begin{proof}
Considering Theorem~1, $\forall 0<\lambda_0<0.5$, $\therefore \lambda_0<\lambda_1<0.5$.
For next iteration, $\because 0<\lambda_0<\lambda_1<0.5$, $\therefore 0<\lambda_1<\lambda_2<0.5$.
Thus, it can be easily deduced that,
\begin{equation}
\forall n\in \mathbb{N} \& n\ge 2, 0<\lambda_{n-1}<\lambda_n<0.5
\label{eq:ap1}
\end{equation}

And for each iteration, the increment $\Delta(\lambda_{n}) = \lambda_{n}-\lambda_{n-1}=\lambda_{n-1}-2\lambda_{n-1}^2$.
Because $0<\lambda_{n-1}<\lambda_n<0.5$, therefore 
\begin{equation}
0<\Delta(\lambda_n)\le 0.125
\label{eq:ap2}
\end{equation}. 

We can use proof by contradiction here.
Suppose $\lim\limits_{n\rightarrow\infty}\lambda_n=\Lambda$, where $\Lambda\ne 0.5$.
Then, according to Eq.~\ref{eq:ap1}, $\Lambda$ must be $<0.5$.
When $n\rightarrow\infty$, $\Delta(\Lambda)$ has to be $\rightarrow 0$.
But this is contradicted to (Eq.~\ref{eq:ap2}).

Therefore, with $n\rightarrow \infty$ (large enough), $\lim\limits_{n\rightarrow\infty}\lambda_n=0.5$.
\end{proof}

Note that the reset operation before peudo-read in accurate $[0,1]$ RNG is to guarantee $0<\lambda_0=p_{BFR}\le 0.5$.

\section{Appendix: List of All Variables}\label{sec:appx2}

\begin{tabular}{{m{0.2\linewidth} | m{0.7\linewidth}}}
	\hline
	\hline
	\textbf{Variable} & \textbf{Description} \\
	\hline
	\centering $p(x)$&  the prior distribution \\
	\hline
	\centering$q(x)$ &  the transfer distribution \\
	\hline
	\centering$t_{pr}$ &  time to trigger the random bit flip phenomenon \\
	\hline
	\centering$X$ &  set of all possible sampling values \\
	\hline
	\centering$x^{*}$ &  newly sampled candidate value \\
	\hline
	\centering$x^{i}$ &   value accepted and stored in the previous address\\
	\hline
	\centering$x^{i+1}$ &  value accepted and stored in the present address \\
	\hline
	\centering$A_{start}$ &  the starting address for the target \\
	\hline
	\centering$A_{end}$ &  the ending address for the target \\
	\hline
	\centering$A$ &  present address \\
	\hline
	\centering$u$ &  random number used in accept ratio calculation\\
	\hline
	\centering$\alpha$ & the accept ratio \\
	\hline
	\centering$p_{BFR}$ & bit flip rate of SRAM bitcells \\
	\hline
	\centering$\lambda$ & probability of bit as 1 in a generated multi-bit sample\\
	\hline
	\centering $R_n^l$ & the $l^{\textrm{th}}$ generated sample in the $n^{\textrm{th}}$ stage in the multi-stage XOR gate scheme \\
	\hline
	\hline
\end{tabular}